# FUTURE OF SUPPLY CHAIN: CHALLENGES, TRENDS, AND PROSPECTS

Cristiana L. Lara[1*], John Wassick[2]
[1]Modeling and Optimization, Amazon.com, Bellevue WA 98004
[2]Department of Chemical Engineering, Carnegie Mellon University, Pittsburgh PA 15213

*Abstract*

This paper discusses the broad challenges shared by e-commerce and the process industries operating global supply chains. Specifically, we discuss how process industries and e-commerce differ in many aspects but have similar challenges ahead of them in order to remain competitive, keep up with the always increasing requirements of the customers and stakeholders, and gain efficiency. While both industries have been early adopters of decision support tools based on machine intelligence, both share unresolved challenges related to scalability, integration of decision-making over different time horizons (e.g. strategic, tactical and execution-level decisions) and across internal business units, and orchestration of human and computer-based decision-makers. We discuss future trends and research opportunities in the area of supply chain, and suggest that the methods of multi-agent systems supported by rigorous treatment of human decision-making in combination with machine intelligence is a great contender to address these critical challenges.

**Introduction**

The concept of supply chain as an interconnected, dynamic physical system to deliver goods and services to consumers has become a popular topic for the public in recent years. This increased awareness is due to the recognition by the public of the connection between supply chain problems and the lived experience of disappoints concerning product availability and price. Past globalization of trade, and the enterprises that engage in it, has increased their scope and their complexity. Consequently, supply chains are more challenging to operate and more vulnerable to disruptions. The recent pandemic, high profile logistical disruptions, and war have exposed the vulnerability of global supply chains. The corresponding swings in consumer preferences and demand have stressed the management practices surrounding supply chain operations. Both companies serving end consumers and those farther upstream have been focused on modernizing and digitizing their supply chains to retain and grow their customer base and their bottom line.

The ongoing activities by the many diverse operating companies pursuing improvements to their supply chains raises the question of shared ideals for their future supply chains. It is safe to say that all companies seek to be more flexible and agile in the face of disruptions, more adaptable to evolving markets, and more aligned to suppliers and customers in general. However, they also share specific objectives at an operational level. This can be demonstrated by examining the shared challenges of supply chains of e-commerce and the process industries. Both are challenged with the scalability and integration needed to deliver end-to-end performance. Both operate enterprise scale

---

workflows that require coordination of decision-making over different time granularities and geographic areas, i.e., closing the gap between long-term planning, tactical planning, and execution. Both are deploying advanced analytical methods to work in concert with the human decision makers. Considered together, the aspirations for supply chain in e-commerce and the process industries provides a consistent view of the future of supply chain and research to support it.

This paper is structured as follows. The difference in the nature of the supply chains for the process industry and e-commerce are first described. Their shared challenges are discussed next. Then future trends and research opportunities are identified to address the shared challenges. Finally, in a future where artificial intelligence is pervasively used in supply chain operations, the place for the use of human intelligence is indicated.

**Different Industries, similar challenges**

*Process Industries*

Within the broad process industries, the chemical industry operates some of the most complex and challenging supply chains. Key reasons include level of vertical integration of product lines, diversity of markets served and position in value chains, energy intensity, and capital intensity. Large chemical companies have complex, globally distributed manufacturing sites that are themselves tightly coupled, internal supply chains (Wassick, 2009).

Recent figures from The Dow Chemical Company provide a good example of the scale and complexity of a global chemical company (*SupplyChainWorld* 2015). Manufacturing more than 6,000 product families at 201 sites in 35 countries, Dow manages roughly 6,000 shipments per day to serve 45,000 customer locations. The company's annual shipment volume is about 130 billion pounds and uses 450 warehouses and 150 contract terminals. The company works with 650 service providers in 160 countries and receives more than 110 billion pounds of raw materials and intermediates each year from 4,000 suppliers.

We will use the Supply Chain Operations Reference (SCOR) model, commonly used by supply chain professionals as a paradigm for discussing supply chains, and comprised of the business processes *Plan*, *Source*, *Make*, *Deliver*, and *Return*, to briefly discuss the nature of chemical industry supply chains (Delipinar and Kocaoglu, 2016).

Central to the *Plan* process are strategic and tactical supply and demand balancing, strategies to support a business financial plan, and long-term asset strategy. These decisions become very challenging when vertical integration requires planning across multiple businesses, each with its own profit and loss metric, serving diverse markets. Business supply chains are highly interdependent with each echelon serving both external markets and internal downstream businesses. Inventory management is confounded by the various dimensions of the value of upstream inventory.

The *Source* process, responsible for raw material acquisition and the associated supplier management, has several complexities in the chemical industry. For one, external sources can be used to supplement the internal sources found in the vertical integration of a company. Raw materials can include globally traded commodities like oil, naphtha, and natural gas for which long term hedging and contracts can be used for procurement while also exploiting spot market prices. Large manufacturing sites can operate cogeneration electrical power plants that serve the onsite production units but also can supply to the local electrical grid, or power can be purchased from the grid when advantageous.

The *Make* process focuses on the broad set of manufacturing activities which includes production, production scheduling, packaging, and site logistics. Safety and environmental protection are top priority in the chemical industry and the nature of the products and processes that make them leads to complex business rules and constraints. Here too, vertical integration at a manufacturing site adds another layer of constraints. Logistics operations can include massive railyards, tank farms, and marine ports.

Within the *Deliver* process are activities associated with order management, warehousing, and transportation. Most of the commerce in the chemical industry is made up of business-to-business transactions which can involve long lead time orders of massive volumes and complicated sales contracts. Transportation for specific orders may be supplied by ocean going vessels, barges, railcars, and dedicated tank trucks, and combinations of these. International shipments to customers are common which requires sophisticated international trade operations to manage compliance with duties and tariffs and shipping requirements.

*E-commerce*

Online retailers have come of age with the dot-com boom, and since then, have been competing with other retailers within the broader retailing industry. Xu (2005) summarizes some characteristics of online retailing, reinterpreted in the list below:

- *Large Selection:* Not limited by physical space in the store front, online retailers are able to offer hundreds of thousands of Stock-Keeping Units (SKU) in stock.
- *Logistics as a Matter of Trust:* Online customers consider the timely delivery of products to be a significant component of trust. Thus, the reliability and efficiency of the supply chain are even more crucial than for traditional *brick-and-mortar* retailers.
- *Data-driven decision making:* These companies rely on rigorous data analytics to make decisions at every stage

of supply chain, from selection of products and inventory placement to delivery.
- *Flexibility when fulfilling customer orders:* In online retailing, there is typically a time delay between when a demand occurs (customer order) and when inventory is physically deployed and transported to meet a customer order. By delaying the decisions on how to fulfill customer orders, these companies can make better operational decisions to utilize their resources and information more efficiently. For instance, e-tailers can profit from bundling multiple items ordered by the same customer into a single shipment. By moving products after orders are placed, an online retailer has flexibility to decide from which of its warehouses or fulfillment centers it is going to serve a customer demand, providing new opportunities to minimize operating costs (Janis, 2019).

E-commerce supply chain structure can be divided into three main pieces: *first-mile*, *middle-mile* and *last-mile*:
- *First-mile* refers to the transportation of goods across the first leg of the supply chain, which in the case of e-commerce is the inbound transportation between vendors/merchants and the distribution centers or fulfillment center. The inbound network has global scale, uses a combination of transportation modes (e.g., ships, airplanes and trucks), and deals with time horizon of the order of weeks to months. The main technical challenges in the *first-mile* space are related to inbound network design, and inventory placement and replenishment (Chen, 2017, Govindarajan et al, 2020, DeValve et al, 2021).
- *Middle-mile* encompasses the outbound transportation network between the fulfillment center and the delivery nodes. The outbound network has national/continental scale, uses a combination of transportation modes (e.g., trucks and airplanes). As customers got accustomed to faster delivery timelines (from days to hours), new challenges have arisen in the *middle-mile* space to be able to design a transportation network that allows aggressive delivery promises while keeping operations cost-effective (Lara et al, 2022).
- *Last-mile* is the last leg of the supply chain to deliver packages to customers' doors. It accounts for a significant part of the total fulfillment cost, but deals with much smaller scales: it focuses on city/county level, and the time-horizon is of the order of hours. The main technical challenge in this space is related to vehicle routing (Delling et al, 2017).

Traditionally, each of these steps was performed by different companies: e-commerce companies would rely on third party logistics to transport between vendors and the fulfillment center and between the fulfillment center and the final customers. However, more recently, some players have invested in verticalization as a way to be less susceptive of global disruptions such as the 2021 supply chain crisis (Schoolov, 2021).

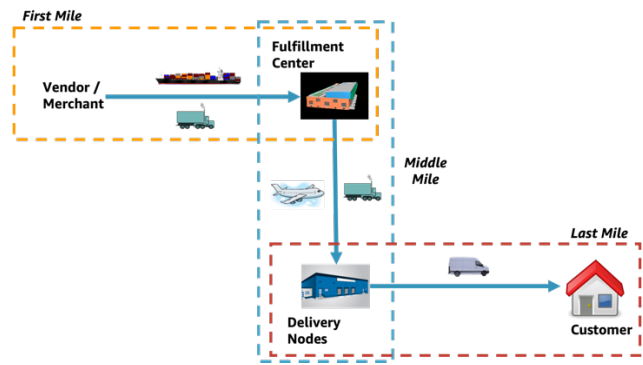

*Figure 1. E-commerce supply chain structure*

To exemplify the complexity of e-commerce operations and decisions, following are some recent figures from Amazon.com. Their logistics network includes more than 110 aircrafts, 50,000 trailers, 400 fulfillment centers, 150 sortation centers, and 1,000 delivery stations globally (Amazon, 2022), to store and transport hundreds of millions of products sold across multiple geographies (Amazon, 2021).

Using the SCOR model as reference again, the main difference between e-commerce and the process industry lies in the *Source* and *Make* processes. E-commerce retailers buy ready-to-sell goods from vendors; hence, the manufacturing of the product is not included as part of their supply chain. The *Source* process becomes responsible for demand forecasting and inventory planning. The *Make* process skips production and focuses on packaging and site logistics. For e-commerce, the *Deliver* process is the core of the business, i.e., fulfilling customer orders and delivering to their home.

*Challenges*

While the Process and E-commerce industries differ in structure, end-customer and overall processes, they share similar issues, and have similar challenges to overcome in order to remain competitive, keep up with the always increasing requirements of the customers and stakeholders, and move towards a more digitalized and efficient supply chain.

It is well understood that analytical methods support a mathematically rigorous description of the problem to be solved, which provides for precise definition of the role of the variables involved, the analysis of the current situation, and numerical assessment of the alternative actions. Thus, analytical methods enforce a discipline and consistency to decision-making that is not guaranteed by humans, but which is generally sought. Efficiency is also gained as computers can execute the decision steps much faster than humans. The decision domain is also expanded as analytical methods are capable of handling problems whose scope and complexity exceeds the capacity of human intelligence. Thus, the application of the strengths of machine intelligence to supply chain decision-making has been the subject of a tremendous amount of research and the basis of sophisticated commercial software (Wenzel, 2019, Ni, 2020).

The process industries use many advanced analytical methods, embodied in commercial software, in supply chain operations. Optimal design or redesign of a supply chain network using mathematical optimization has gained wide acceptance and is routinely practiced (supports *Plan*) (Bassett, 2018). Realtime optimization of world scale process plants based on the solution of flowsheet models has widespread applications (supports *Make*) (AspenTech, 2022). AI powered predictive analytics is providing real time transportation visibility to a variety of industries, including the process industry (supports *Deliver*) (FourKites, 2021). For problems not suitably addressed by commercial offerings, operating companies have been active in developing inhouse solutions based on advanced methods. For the most part, success with machine intelligence has occurred at specific decision points in the enterprise scale workflow of supply chain operations. This leaves integration gaps that need to be addressed both horizontally and vertically.

Similarly, e-commerce companies heavily rely on machine learning, mathematical optimization and simulation to plan, manage and execute their supply chains. For example, at Amazon, forecasting models predict the demand for every product, buying systems determine the right level of product to purchase from different suppliers, while large-scale placement systems determine the optimal location for products across hundreds of fulfillment centers (Amazon, 2021). Inside the fulfillment centers, computer vision systems keep track of where every product is. When an order is placed, it triggers multiple systems that have to solve large-scale optimization problems in real time to decide how to pick, pack, load, and move packages to the customer's house (Karlinsky, 2019). The real-time decisions rely on previously made tactical (e.g., truck, flight and labor scheduling) and strategic decisions (e.g., where to build new fulfillment centers, sortation center and delivery stations and for what capacity) that are also supported by analytical tools.

While there have been significant advancements in computing power (Hao, 2019), machine learning (Stanford University, 2021), and optimization solvers (Bixby, 2012), the size and complexity of supply chains grew in even faster pace, making it intractable to optimize the entire network and operations as a single problem. The question becomes: how to best break down these problems in a way that makes business coordination easier, and how to use modeling to help coordinate between multiple organizations and reconcile between their conflicting objectives.

This problem decomposition in the modeling space tends to follow the same structure as the business division in the organizational space. And as it happens in the modeling space, it becomes challenging to make sure that all organizations use a consistent set of data and assumptions and are optimizing for the overall business (global optimal) instead of their own specific metrics (local optimal). In the case of e-commerce, it seems intuitive to break the problem down between first-mile, middle-mile and last-mile, as they deal with different scales and tackle different parts of the network. However, the decisions made in each problem deeply impacts the others. For example, the inventory placement decision (first-mile) is critical for the design of the middle-mile network. If one optimizes for the first-mile alone, it may seem like a good idea to have a huge fulfillment center that stores all products and is close to a shipping port. The inbound transportation cost to the fulfillment center is minimal, and this structure eliminates the complexity of trying to decide which fulfillment center should carry which product. However, such solution significantly hurts the middle-mile. Most orders would have to travel long distances between the fulfillment center and customers' homes, increasing outbound transportation cost, hurting delivery speed and, in turn, customer satisfaction.

Another crucial challenge is how to better coordinate decision-making across different time scales. Traditionally, long-term strategic decisions, mid-term tactical decisions and short-term execution decisions are made by different teams. The closer the timeline the more information available but the less flexibility for significant modifications to the plan. There is a large body of literature on how to coordinate between strategic, tactical and execution-level decisions (Brunaud, 2019), but in practice there is still a lot to be done in this area to reduce the gap between plan and execution. Planning should be better at taking variability and uncertainty into account so it can actually be executed, and the real-time systems should be better at adhering to the original plan.

Additionally, data issues are often a significant barrier to the successful application of advanced analytical methods. The data involved in an enterprise workflow such as supply chain operations spans many diverse data repositories which are not well connected for data exchange, lack consistency in naming conventions, and have poorly defined data relationships. To be able to have more automated processes and take full advantage of computer-based analytical processes, companies need to ensure that: (i) everyone is making the same assumptions and using the same source of truth for data; (ii) data can be automatically aggregated/disaggregated to support consistency between multiple time scales; and (iii) data relationships can be modeled to support end-to-end process and business coordination.

Finally, the role of the human intelligence in supply chain decisions needs more rigorous treatment. With very few exceptions, advanced analytical methods applied in supply chain operations rely on some form of human intervention or assistance in carrying out the intended decision-making role. It is important to understand where human reasoning is more effective versus computer-based analytical methods and how to blend both cognitive approaches along decision points in the end-to-end process for the most effective solution.

**Future trends and opportunities**

*Future Supply Chain Management*

In the near future, decisions of supply chain professionals will be supported by virtual assistants facilitating the use of advanced analytical methods. Alternative courses of action, derived from an understanding of event propagation or the correlation of multiple events across the supply chain through time, will be provided in a manner that fits within a decision maker's personal workflow. Logic and machine learning will derive insights from trends, identify deviations from normal behavior, or derive other patterns that signify or predict actionable business interventions such as opportunities to reduce costs or working capital. Supply chain professionals at all levels will have quantitative, financial measures to assess the alternatives involved in their decisions, providing them with complete visibility of the impact of decisions on financial results.

In the long-term, there is ever increasing scope of decision-making automation and a diminishing reliance on manual intervention. Reminiscent of the level of automation achieved in manufacturing, automation of short-term, operational level supply chain decisions will be commonplace. Supply chain professionals will manage the process of their role as carried out by one or more intelligent software agents, rather than exercise direct decision-making. Thus, decision makers are released from day-to-day decision-making to focus on more strategic decisions and value-added activities.

This future system of agents, both human and computer-based, will be supported by a common supply chain digital twin tightly aligning the decision processes of all decision-makers. The digital twin spanning the entire supply chain, and fed by operational, transactional, and cash flow data, will allow the system to operate with fully integrated strategic, tactical, and operational decisions. There is no siloed decision making. Myopic and chaotic decisions are avoided as decision makers have full upstream and downstream awareness of decision implications. The resulting digital supply chain will operate with competitively advantaged agility and adaptability while balancing cost and other financial measures across the supply chain to achieve alignment among all stakeholders.

*Promising Research Directions*

The properties associated with the human dimension of supply chain decisions and the multi-agent nature of supply chain operations have not received widespread attention in the PSE literature. We believe both these streams of research will have profound impact on the challenges of scalability, business coordination, multiple time scales, and the prevalence of human decision-making.

Research on design of analytical methods that rigorously account for human input, response, and intervention is needed to balance the tremendous amount of current research developing model-based solutions. The emerging area of Cognitive Engineering is developing mathematical models that describe human decision-making, as well as analytical methods that incorporate human intuition and judgement. Engineering human-in-the-loop interactions as carried out in cyber-physical systems is another good source of ideas for supply chain operations (Gil et al, 2020). Developing machine learning methods to extract from historical records policies practiced by humans is another fruitful area of research. Ultimately PSE research should lead to holistic solution designs where mathematical model is just one component and hybrid intelligence is achieved.

The field of multi-agent systems provides the foundation for achieving the future supply chain described in the previous section (Jones, 2018). As a collection of multiple decision-making agents which interact in a shared environment to achieve common or conflicting goals, a multi-agent system is an excellent paradigm to analyze and synthesize an end-to-end process like supply chain. The definition of an agent and the structure of the system are highly flexible. Human or artificial agents are accommodated, robotic warehouses and autonomous trucks/ships can be modeled as agents, and agents can reflect existing organizational roles, or organizations themselves. Multi-agent systems perform distributed decision-making offering a natural approach to decomposing problems that are intractable when taken as a whole. Traditional decomposition methods used in optimization can be viewed as multi-agent systems. Because agent connections are flexible and dynamic to create solutions as needs arise, multi-agent systems are naturally more agile than hierarchical methods, and therefore more resilient.

*Final Thoughts*

One motivation for our proposed research directions is that we agree with other researchers on the complimentary nature of human intelligence and machine intelligence (human-AI symbiosis) (Jarrahi, 2018). We consider machine intelligence to include artificial intelligence and machine learning, as well as the broad spectrum of traditional process systems engineering methods like optimization modeling, simulation modeling, and general regression modeling. Given the strengths of these methods as cited earlier, machine intelligence will clearly be central to the future supply chain.

However, the operation of massive and evolving supply chains as found in e-commerce and the process industries involves problems where human intelligence has advantages. The external connections found in supply chains create decision domains with limited information for which human intuition is well suited. Machine intelligence can produce spurious results or impractical solutions and can be subjected to situations where assumptions do not hold. Under these conditions, common sense practiced by humans is important in reaching effective decisions. Human creativity provides solutions to problems not previously encountered such as occur in supply chain disruptions. Supply chain networks are also social networks comprised of suppliers and customers, enterprise-scale workflows involving many employees, and a variety of external stakeholders. Humans, as social creatures, have the natural ability to account for the tendencies, priorities, motivations of others and the tendency to engage with others for mutual success. Notwithstanding the tremendous progress of machine intelligence, we see human intelligence as critical to the success of the future supply chain. When both are used to their greatest advantage the future supply chain will meet the challenges of today, adapt to the challenges of the future, satisfy the needs of all stakeholders, and recede from the headlines.

**Conclusions**

Modern, global industries, such as e-commerce and chemical manufacturing, operate massive supply chains with complex enterprise-scale workflows. While both industries have been early adopters of decision support tools based on machine intelligence, both share unresolved challenges, such as those related to the scale of their supply chain operations, the integration of strategic, tactical, and operational decisions, harmonizing across internal business units, and the effective orchestration of the many decision-makers, both human and computer-based. We believe the methods of multi-agent systems supported by rigorous treatment of human decision-making has excellent potential to address these critical challenges and complement the existing contributions of the Process Systems Engineering community.

**References**

Amazon (2021). The evolution of Amazon's inventory planning system. *Amazon Science*, October 1, 2021. https://www.amazon.science/latest-news/the-evolution-of-amazons-inventory-planning-system

Amazon (2022). Amazon is ready to deliver for customers and sellers this Prime Day. *Amazon News*, July 5, 2022. https://www.aboutamazon.com/news/transportation/amazon-is-ready-to-deliver-for-customers-and-sellers-this-prime-day

AspenTech (2022) https://www.aspentech.com/en/products/msc/aspen-gdot

Bassett, M. (2018) Optimizing the Design of New and Existing Supply Chains at Dow AgroSciences. *Computers & Chemical Engineering* 114:191-200.

Bixby, R. E. (2012). A brief history of linear and mixed-integer programming computation. *Documenta Mathematica*. 107-121.

Brunaud, B. (2019) Models and Algorithms for Multilevel Supply Chain Optimization, Ph.D. thesis, Carnegie Mellon University, Pittsburgh, PA.

Chen, A.I. (2017). Large-scale Optimization in Online-retail Inventory Management, Ph.D. thesis, MIT, Cambridge, MA.

Delipinar, G.E., Kocaoglu, B. (2016) Using SCOR Model to Gain Competitive Advantage: A Literature Review. *Procedia - Social and Behavioral Sciences*. 229: 398-406.

Delling, D., Goldberg, A.V., Pajor, T., Werneck, R.F. (2017) Customizable Route Planning in Road Networks. *Transportation Science* 51(2):566-591.

DeValve, L., Wei, Y, Wu, D., Yuan, R. (2021) Understanding the Value of Fulfillment Flexibility in an Online Retailing Environment. *Manufacturing & Service Operations Management* 0(0).

FourKites (2022) https://www.fourkites.com/platform/real-time-visibility/

Gil, M., Albert, M., Fons, J., Pelechano, V. (2020) Engineering human-in-the-loop interactions in cyber-physical systems. *Information and Software Technology* 126: 106349.

Govindarajan, A, Sinha, A, Uichanco, J. (2021) Joint inventory and fulfillment decisions for omnichannel retail networks. *Naval Research Logistics*. 68: 779-794.

Hao, K. The computing power needed to train AI is now rising seven times faster than ever before, November 11, 2019. *MIT Technology Review*. https://www.technologyreview.com/2019/11/11/132004/the-computing-power-needed-to-train-ai-is-now-rising-seven-times-faster-than-ever-before/

Jasin, S., Sinha, A., Uichanco, J. (2019). Omnichannel Operations: Challenges, Opportunities, and Models. In: Gallino, S., Moreno, A. (eds) Operations in an Omnichannel World. Springer Series in Supply Chain Management, vol 8. Springer, Cham. https://doi.org/10.1007/978-3-030-20119-7_2

Jarrahi, M.H. (2018) Artificial intelligence and the future of work: Human-AI symbiosis in organizational decision making. *Business Horizons* 61(4): 577-586.

Jones, A.T., Romero, D., Wuest, T. (2018) Modeling agents as joint cognitive systems in smart manufacturing systems. *Manufacturing Letters* 17:6-8.

Karlinsky, N. (2019). How artificial intelligence helps Amazon deliver. *Amazon News*, June 5, 2019. https://www.aboutamazon.com/news/innovation-at-amazon/how-artificial-intelligence-helps-amazon-deliver

Lara, C.L., Koenemann, J., Nie, Y., de Souza, C.C. (2022), Scalable Timing-Aware Network Design via Lagrangian Decomposition. *Submitted for publication*.

Ni, D., Xiao, Z. & Lim, M.K. (2020), A systematic review of the research trends of machine learning in supply chain management. *Int. J. Mach. Learn. & Cyber*. 11, 1463–1482.

Schoolov, K. (2021). Amazon Is Making Its Own Containers and Bypassing Supply Chain Chaos with Chartered Ships and Long-Haul Planes." *CNBC*, December 4, 2021. https://www.cnbc.com/2021/12/04/how-amazon-beats-supply-chain-chaos-with-ships-and-long-haul-planes.html.

Stanford University. Gathering Strength, Gathering Storms: The One Hundred Year Study on Artificial Intelligence (AI100) 2021 Study Panel Report. *Stanford University*, Stanford, CA, September 2021. http://ai100.stanford.edu/2021-report.

SupplyChainWorld (2015). The Dow Chemical Company. *SupplyChainWorld*. https://scw-mag.com/profiles/536-the-dow-chemical-company/

US Department of Commerce Retail Indicator Division (2022). Quarterly Retail E-Commerce Sales 4th Quarter 2021. United States Census Bureau.

Wassick, J.M. (2009) Enterprise-wide optimization in an integrated chemical complex. *Computers & Chemical Engineering* 33(12):1950-1963.

Wenzel, Hannah; Smit, Daniel; Sardesai, Saskia (2019) : A literature review on machine learning in supply chain management, In: Kersten, Wolfgang Blecker, Thorsten Ringle, Christian M. (Ed.): Artificial Intelligence and Digital Transformation in Supply Chain Management: Innovative Approaches for Supply Chains. Proceedings of the Hamburg International Conference of Logistics (HICL), Vol. 27, ISBN 978-3-7502-4947-9, epubli GmbH, Berlin, pp. 413-441.

Xu, P. (2005). Order Fulfillment in online retailing: what goes where. Ph.D. thesis, MIT, Cambridge, MA.